\documentclass[lettersize,journal]{IEEEtran}
\IEEEpubidadjcol
\usepackage{amsmath,amsfonts}
\usepackage{algorithmic}
\usepackage{algorithm}
\usepackage{array}
\usepackage[caption=false,font=normalsize,labelfont=sf,textfont=sf]{subfig}
\usepackage{textcomp}
\usepackage{stfloats}
\usepackage{url}
\usepackage{verbatim}
\usepackage{graphicx}
\usepackage{cite}
\usepackage{multirow}
\usepackage{booktabs}
\usepackage{newtxmath}
\usepackage{float}
\usepackage{xcolor}
\begin{document}

\title{CoAVT: A Cognition-Inspired Unified Audio-Visual-Text Pre-Training Model for Multimodal Processing}

\author{Xianghu Yue$^{*}$, Xiaohai Tian, Lu Lu, Malu Zhang, Zhizheng Wu, Haizhou Li,~\IEEEmembership{~IEEE Fellow}
\thanks{$^{*}$This work was done when Xianghu Yue was an intern in ByteDance.}
}

\markboth{Journal of \LaTeX\ Class Files,~Vol.~14, No.~8, August~2021}%
{Shell \MakeLowercase{\textit{et al.}}: A Sample Article Using IEEEtran.cls for IEEE Journals}

\IEEEpubid{0000--0000/00\$00.00~\copyright~2021 IEEE}

\maketitle

\begin{abstract}
There has been a long-standing quest for a unified audio-visual-text model to enable various multimodal understanding tasks, which mimics the listening, seeing and reading process of human beings.
Humans tends to represent knowledge using two separate systems: one for representing verbal (textual) information and one for representing non-verbal (visual and auditory) information. 
These two systems can operate independently but can also interact with each other.
Motivated by this understanding of human cognition, in this paper, we introduce CoAVT -- a novel cognition-inspired Correlated Audio-Visual-Text pre-training model to connect the three modalities.
It contains a joint audio-visual encoder that learns to encode audio-visual synchronization information together with the audio and visual content for non-verbal information, and a text encoder to handle textual input for verbal information.
To bridge the gap between modalities, CoAVT employs a query encoder, which contains a set of learnable query embeddings, and extracts the most informative audiovisual features of the corresponding text.
Additionally, to leverage the correspondences between audio and vision with language respectively, we also establish the audio-text and visual-text bi-modal alignments upon the foundational audiovisual-text tri-modal alignment to enhance the multimodal representation learning.
Finally, we jointly optimize CoAVT model with three multimodal objectives: contrastive loss, matching loss and language modeling loss.
Extensive experiments show that CoAVT can learn strong multimodal correlations and be generalized to various downstream tasks.
CoAVT establishes new state-of-the-art performance on text-video retrieval task on AudioCaps for both zero-shot and fine-tuning settings, audio-visual event classification and audio-visual retrieval tasks on AudioSet and VGGSound.
The results demonstrate the effectiveness and
superiority of the proposed model for multimodal processing.
\end{abstract}

\begin{IEEEkeywords}
Multi-modal pretrain, representation learning, contrastive learning
\end{IEEEkeywords}

\section{Introduction}
\IEEEPARstart{H}{umans} learn by reading, seeing and listening, which is a typical process of multimodal processing involving text, visual and audio content, and use the acquired knowledge to understand and interact with the world.
Multimodal processing~\cite{clip4vla, vatt, valor, mcn, clip}, which aims to learn the general knowledge across multiple modalities, has obtained much attention recently especially with the success of pretraining.
Due to the high complexity and high training cost of multimodal models, most works focus on the processing of two modalities such as text and vision or text and audio. Just like humans benefiting from tri-modal content and their interactions between the modalities,  multimodal understanding tasks rely on an effective tri-modal modeling.

In recent years, pre-training has witnessed a rapid development in multimodal processing, especially for two modalities.
For example, Visual-Language (VL) pre-training models~\cite{actbert, videobert, clip, alpro, blip, blip2}, have shown superior performance on various text-video downstream tasks, such as text-video retrieval and video captioning.
Similarly, Audio-Language (AL) pre-training models, like CLAP~\cite{clap} and LAION~\cite{laion}, have capability to develop audio representation via contrastive learning~\cite{clip} by combining audio data with natural language descriptions.
Based on these bi-modal pre-training methods, we therefore take one step further to learn the general knowledge across three modalities of our daily perception, including audio, visual and text.

Building a unified audio-visual-text model capable of solving various multimodal understanding tasks is a long-standing challenge for multimodal processing research.
Some recent works~\cite{clip4vla, valor, audioclip, opt} attempts to incorporate audio modality into VL pre-training for tri-modal understanding.
However, a common approach in these efforts involves the utilization of three separate encoders for audio, vision and text, and then train it with pair-wise contrastive pre-text tasks.
Although effective, they ignore the inherent alignment between audio and video modalities.
We note that audio and video are two naturally time-aligned and closely related modalities of human perception~\cite{mavil, cavmae}, offering different but complementary information.
With separate and modality-dependent encoders, the synchronization information between audio and video may not be harnessed.
A few recent works~\cite{videcc} employ a dual stream model (one stream being an audio-visual encoder and one stream being a text encoder) and solely train it with audiovisual-text contrastive loss for coarse-grained alignment.
However, the modality gap between the three distinct modalities still exists and the inherent bi-modal correspondences (e.g., audio-text and visual-text) are not fully exploited.

\IEEEpubidadjcol

\begin{figure}
\centering
\includegraphics[scale=0.65]{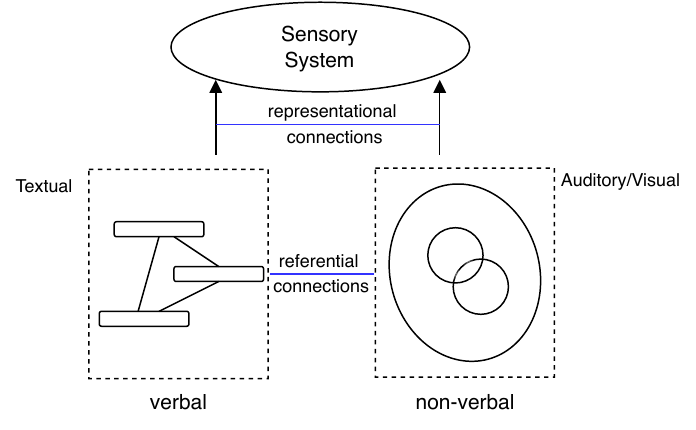}
\caption{The dual coding theory of human cognition proposed by Paivio~\cite{dual1}.}
\label{dual_coding}
\end{figure}

When compared to machines, the human brain has an extraordinary ability to perceive and process multimodal information~\cite{dual1, dual2}.
Human cognition process is a useful tool of reference for machine multimodal representation learning.
According to the dual coding theory~\cite{dual1}, as illustrated in Figure~\ref{dual_coding}, human cognition is unique in that it has become specialized for dealing simultaneously with language and with non-verbal objects and events.
The theory assumes that there are two cognitive subsystems, one specialized for the representation and processing of non-verbal objects and events (i.e., auditory and imagery), and the other specialized for dealing with language.
Moreover, the dual coding theory identifies two types of connections, one is the representational connection which represents the direct activation of verbal and non-verbal representations, and the other is the referential connection which represents the activation of the verbal system by the non-verbal system or vice-versa. 

Inspired by human cognition mechanisms, this study proposes a cognition-inspired unified audio-visual-text pre-training model, namely Correlated Audio-Visual-Text pre-training (CoAVT), to learn multimodal representations for solving various multimodal understanding tasks.
CoAVT aims to build connections between different modalities and learn rich multimodal representation.
Specifically, a joint audio-visual encoder is employed to handle audio and visual input simultaneously, thus leveraging the natural alignment between audio and video, and a text encoder for textual input.
The joint audio-visual encoder is designed for non-verbal information, while the text encoder is for verbal information, similar to the two cognitive subsystems of human.
To mitigate the modality gap, we propose a query encoder as a bridge between the joint audio-visual encoder and the text encoder.
The query encoder contains a set of trainable query embeddings to interact with the joint audio-visual encoder and extract the most informative audiovisual representation of the corresponding text.
Furthermore, to leverage the correspondences between audio and vision with language, we further build the audio-text and visual-text bi-modal alignments upon the foundational audiovisual-text tri-modal alignment simultaneously to enhance the multimodal representation learning.
Finally, besides the simple contrastive loss for coarse-grained alignment, we jointly optimize CoAVT with matching loss for fine-grained alignment and causal language modeling loss for contextual coherence.

Extensive experiments are conducted on multiple downstream tasks, including text-video retrieval with different experimental setups (i.e. zero-shot and fine-tune), audio-visual event classification and audio-visual retrieval.
The results demonstrate that our proposed CoAVT model is able to learn better cross-modal alignment, and consistently outperforms current SOTAs on the benchmark datasets of AudioCaps, AudioSet and VGGSound.
It achieves an average performance improvement of 12.4\% Recall@10 score and 1.8\% Recall@1 score on the zero-shot and fine-tune setting of the retrieval datasets respectively, while the classification accuracy improvement is 2.5\% mAP.

In summary, we make the following contributions:
\begin{itemize}
    \item We introduce CoAVT, a cognition-inspired unified pre-training model capable of solving various multimodal understanding tasks across multiple modalities, including audio, vision and text.
    \item To mitigate the modality gap between the three modalities, we introduce the query encoder as a bridge for effective audiovisual-text alignment learning.
    \item To fully exploit the inherent bi-modal correspondences, we build bi-modal audio-text and visual-text alignments upon the foundational audiovisual-text tri-modal alignment, which explicitly correlate the three modalities.
    \item Our model achieves the state-of-the-art performance on text-video retrieval task on AudioCaps for both zero-shot and fine-tuning settings, audio-visual event classification and audio-visual retrieval tasks on AudioSet and VGGSound.
\end{itemize}

The remainder of this article is organized as follows.
Section~\ref{related_work} reviews advances in bi-modal vision/audio-language and tri-modal audio-visual-text pre-training methods. 
Section~\ref{method} describes our proposed method. 
Section~\ref{exps} presents the experimental results and detailed analysis.
The conclusion and future work are presented in Section~\ref{conclusion}.

\section{Related work}
\label{related_work}
In this section, we briefly revisit the bi-modal audio/vision-language pre-training models and tri-modal audio-visual-text pre-training models to set the stage of this work.

\subsection{Vision/Audio-Language Pre-training}
Vision-language pre-training aims to learn bi-modal foundation models with improved performance on various vision-language downstream tasks, such as video retrieval, video question answering, and video captioning.
Most of them~\cite{videobert, cbt, lxmert, actbert, clip, clipbert} focus on the visual-language alignment within the context of videos.
Pioneering works such as VideoBERT~\cite{videobert} and CBT~\cite{cbt} explored the potential of joint visual-language representation via self-supervised learning.
For fine-grained multimodal understanding, HERO~\cite{hero} adopts a hierarchical structure to encoder video and text and employs a temporal-specific pre-text task, while UniVL~\cite{univl} designs a generation pre-text task.
ClipBERT~\cite{clipbert} further introduces an end-to-end manner by inputting sparse sampled frames from video clips rather than densely extracted offline video features from full-length videos.
BLIP-2~\cite{blip2} bootstraps VL pre-training from off-the-shelf frozen pre-trained vision encoders and froze large language models.
Collectively, these works well explore the correlation between vision and text modalities.

In a similar vein, audio-language pre-training~\cite{clap, laion, wavcaps, tap, balt} seeks to establish a profound comprehension of audio content by connecting audio and natural language.
Following the success of CLIP~\cite{clip} that learns image representations with natural language supervision, CLAP~\cite{clap} and LAION~\cite{laion} bring audio and text descriptions into a joint multimodal embedding space through contrastive learning, which are pre-trained on a large amount of audio-text pairs.
BLAT~\cite{balt} takes this a step further, bootstrapping audio-language pre-training with synthetic data.
These works well capture the alignment and learn multimodal representation between audio and text modalities.
Their efficacy has been demonstrated across various audio-language downstream tasks, such as audio retrieval, audio captioning and audio event classification.

Building on these advancements, our CoAVT takes a stride forward, aiming to learn multimodal representation among three modalities, including audio, visual and text.

\subsection{Audio-Visual-Text Pre-training}
With the success of pre-training on two modalities, some recent works~\cite{clip4vla, audioclip, wav2clip} try to incorporate audio modality into existing VL pre-training paradigms to achieve tri-modal understanding. 
For example, CLIP4VLA~\cite{clip4vla} and AudioCLIP~\cite{audioclip} are proposed to extend the vision-language model CLIP~\cite{clip} to accommodate audio modality for vision-language-audio multimodal processing.
Based on these, VALOR~\cite{valor} employs three separate encoders for audio, video, and text and pre-trains with contrastive loss and language modeling loss.
VATT~\cite{vatt} introduces a hierarchical contrastive loss for text-video and video-audio alignment, but it targets at learning single-modality representations instead of improving cross-modality capability.
Different above methods, AVR~\cite{videcc} designs a dual stream-model, one stream being an audio-visual encoder and on stream being a text encoder, which is trained with simple contrastive loss, specifically for the video retrieval task.

Our study builds upon these previous work but introduces two major differences in our dual-stream model.
Firstly, to better handle the modality gap and enhance the referential connections between modalities, we employ a query encoder between the joint audio-visual encoder and the text encoder, to extract the most informative features of the text.
Secondly, AVR model only focuses on the representational connections and ignores the correspondences between audio and vision with language while our CoAVT simultaneously exploits their intrinsic correlation for rich multimodal representation learning.


\begin{figure*}[ht]
\centering
\includegraphics[scale=0.75]{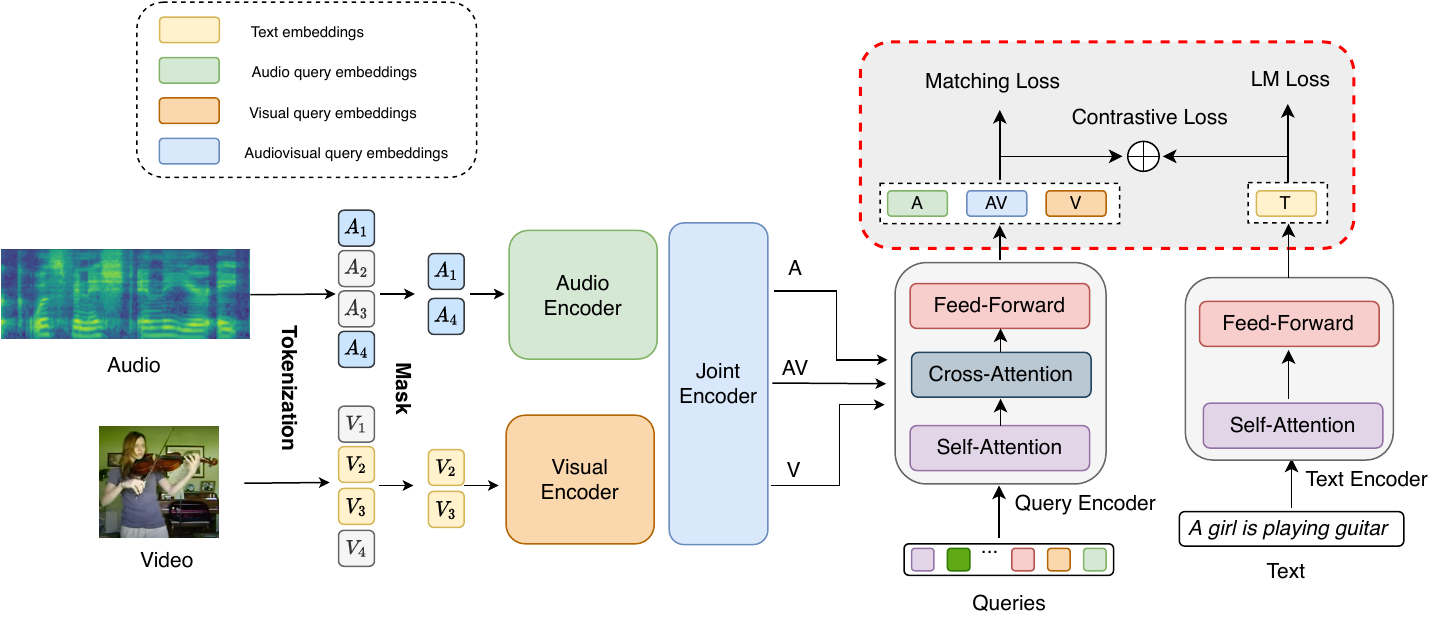}
\caption{The overview of our proposed CoAVT model, which consists of a joint audio-visual encoder, a text encoder and a query encoder, which contains a set of learnable query embeddings.
The query encoder partly shares parameters with text encoder except the cross-attention layers.
The red dashed box shows the pre-training objectives of our CoAVT, which are calculated on three pair-wise losses, including AV-T, A-T, and V-T.
Each pair consists of contrastive loss, matching loss and language modeling loss.
}
\label{fig:framework}
\end{figure*}

\section{Methodology}
\label{method}
In this section, we first introduce important constituent modules of our CoAVT model, we then present the pre-training objectives in detail.
Given a batch of videos and their corresponding descriptions, we first extract the audios from videos and denote the audio batch, video batch and text batch as $A$, $V$, and $T$ respectively.
The goal of our CoAVT model is to effectively capture and exploit the underlying relationships between different modalities (e.g., audio, video and text) and finally to learn rich semantic representations among all three modalities.
By doing so, the corresponding audio, video and text with similar semantics can be embedded close to each other despite being in different modalities, while pushing those with different semantics far away in the multimodal space.

With the multimodal representation fully learned during pre-training, we then fine-tune the model on different downstream tasks including cross-modal retrieval and multimodal event classification to verify the effectiveness.

\subsection{Model Architecture}
Figure~\ref{fig:framework} presents an overview of our CoAVT model, which consists of a joint audio-visual encoder to handle audio and visual signals simultaneously, a text encoder to handle the textual data, and a query encoder to extract audio-visual information that is most informative of the text, thus bridging the modality gap between the three heterogeneous modalities.
Details for each component are as follows.

\subsubsection{Joint Audio-Visual Encoder}
Audio and video are two naturally aligned and closely related modalities of human perception, offering different but complementary information.
Inspired by previous work~\cite{cavmae, mavil, videcc}, our joint audio-visual encoder incorporates an audio-visual encoder pair, and a shallow joint encoder layer.
This design is intended to encode audio-visual synchronization information together with the audio and visual content, promoting enhanced cross-modal understanding and representation learning.

Given a video and its corresponding audio, we follow the pre-processing and tokenization in AST~\cite{ast} and ViT~\cite{vit} for audio and image inputs, respectively.
Specifically, for audio, all audio clips are first randomly cropped or padded to 10 seconds, and then the 128-dimensional log Mel-filterbank features are extracted with a 25ms Hanning window every 10ms, which results in a 1024 $\times$ 128 spectrogram.
We then split the spectrogram into 16 $\times$ 16 square patches $\mathbf{a}=[a^1, a^2, \cdots, a^{512}]$ as the input of the audio encoder.
For video, we uniformly sample 10 RGB frames from each video, and randomly select one frame as input during pre-training. 
For each frame, we resize and center crop it to 224 $\times$ 224 size, and then split it into 16 $\times$ 16 square patches $\mathbf{v}=[v^1, v^2, \cdots, v^{196}]$ as the input of the visual encoder.
The joint encoder employ a multi-stream forward pass strategy, in which we input the output of the audio encoder $E_A^{'}$, the output of the visual encoder $E_v^{'}$, and the concatenated audio-visual representation $[E_A^{'}, E_V^{'}]$ in three independent forward passes and obtain the final audio-only embedding $E_A$, visual-only embedding $E_V$, and joint audio-visual embedding $E_{AV}$.

\subsubsection{Text Encoder}
We use BERT$_{base}$~\cite{bert} as our text encoder.
Given a text input of $N$ tokens, the text encoder outputs an embedding sequence $\{t_{cls}, t_1, \cdots, t_{N}\}$, with $t_i \in \mathbb{R}^d$ and $t_{cls}$ is the embedding of the text [CLS] token.
Following BERT, we choose the $t_{cls}$ as the global text representation.

\subsubsection{Query Encoder}
To mitigate the modality gap, we propose to use the query encoder as a bridge between the joint audio-visual encoder and text encoder, and align the three modalities.
The query encoder inserts one additional cross-attention layer between the self-attention layer and the feed-forward network for each transformer block of the text encoder.
We first create a set number of predetermined learnable query embeddings as input to the query encoder.
The queries can interact with each other through self-attention layers.
More importantly, these queries can further interact with the embeddings (e.g., $E_A, E_V, E_{AV}$) from the joint audio-visual encoder through cross-attention layers, enabling better alignment between modalities.
Except for cross-attention layers, the query encoder shares parameters with the text encoder, thereby these queries can additionally interact with the text through the same self-attention layers as well.
The output of query encoder is $Q \in \mathbb{R}^{N_q \times C}$, where $N_q$ is the number of queries and $C$ is the hidden size.
For the inputs of $E_A, E_V, E_{AV}$, the outputs of query encoder are $Q_A, Q_V, Q_{AV}$, respectively.

\subsection{Pre-training}
In this section, we introduce the pre-training objectives of our CoAVT model in detail.
To conduct unified multimodal representation learning among audio, vision and text, we jointly optimize three objectives during pre-training, including contrastive loss, matching loss and language modeling loss.
For each objective, we mainly consider three modality pairs including text-audio pair (T-A), text-visual pair (T-V), and text-audiovisual pair (T-AV).
Details for each objective are as follows.

\subsubsection{Contrastive Loss}
We first build the coarse-grained alignment between modality $X$ and text via $X$-text contrastive learning, where $X$ represents different modalities including audio (A), visual (V) and joint audio-visual (AV).
It aims to align the embedding space of the query encoder and the text encoder by encouraging positive $X$-text pairs to have similar representations in contrast to the negative pairs.

Formally, we align the output query representation $Q_{X}$ with the text representation $t_{cls}$.
Since $Q_{X}$ contains multiple output embeddings (e.g., 16), we first compute the pairwise similarity between each query output $q_i$ and $t_{cls}$, and then select the highest one as the similarity score:
\begin{equation}
    s(X, T) = max_{i \in N_q} g_{q}(q_i) \cdot g_{t}(t_{cls})
\end{equation}
where $g_q(\cdot)$ and $g_t(\cdot)$ are linear projections that transforms the query embedding and [CLS] embedding to a common normalized low-dimensional space.
The $X$-text contrastive loss consists of two symmetric terms, one for $X$-to-text classification:
\begin{equation}
    \mathcal{L}_{X2T} = -log \frac{exp(s(X_i, T_i)/\tau)}{\sum_{j=1}^{B}exp(s(X_i, T_j)/\tau)}
\end{equation}
and the other for text-to-$X$ classification:
\begin{equation}
    \mathcal{L}_{T2X} = -log \frac{exp(s(T_i, X_i)/\tau)}{\sum_{j=1}^{B}exp(s(T_i, X_j)/\tau)}
\end{equation}
where $\tau$ is a learnable temperature parameter, and $B$ is the batch size. The final X-text contrastive loss is then denoted as:
$\mathcal{L}_{XTC} = \frac{1}{2}(\mathcal{L}_{X2T} + \mathcal{L}_{T2X})$.


\subsubsection{Matching Loss}

$X$-text matching (XTM) is a binary classification task where the model is asked to predict whether $X$-text is matched or unmatched given the corresponding query features and text features.
Here we employ a bi-directional self-attention mask so that all queries and texts can attend to each other, which enables the query encoder to effectively capture  the multimodal information between the text and modality $X$ through the query embeddings.
Finally, we feed each output query embedding into a two-class linear classifier to obtain a matching probability $p_i$, and then average the probabilities across all queries as the final matching score:
\begin{equation}
\mathcal{L}_{XTM} = \frac{\sum_{j=1}^{N_q} ylog p_i }{N_q}
\end{equation}
where $y$ is 1 when the input $X$-text pair is matched and 0 otherwise.


\subsubsection{Language Modeling Loss}
Language modeling aims to generate text given the representations of modality $X$ as the condition.
Specifically, as there exists no direct interactions between the text encoder and the joint audio-visual encoder, the information required for generating the text must be first extracted through the query encoder, and then passed to the text tokens via self-attention layers.
During this process, we employ a multimodal causal self-attention mask to control the interaction between the queries and text, in which the queries can only attend to each other but not the text tokens, and each text token can attend to both all queries and its preceding tokens.
This masking strategy ensures a coherent and effective flow of information from the queries to the text, enabling the generation of contextually relevant and coherent text given the modality $X$ representations.
This casual language modeling loss with modality $X$ as condition can be formulated as:
\begin{equation}
    \mathcal{L}_{XLM} = -\sum_i^{L} log p(y_i|y_{\textless i}, Q_X)
\end{equation}
where $D$ denotes the training batch, $y_i$ denotes the current predicted token, and $y_{\textless i}$ represents the previous predicted tokens. 
The final pre-training objective for modality $X$ is the sum of above three objectives:
\begin{equation}
    \mathcal{L}_{X} = \mathcal{L}_{XTC} + \mathcal{L}_{XTM} + \mathcal{L}_{XLM}
\label{eq6}
\end{equation}

\subsubsection{Overall Pre-training Loss}
Besides the foundational audiovisual-text tri-modal alignment, we further utilize the correspondences between audio and vision with language (e.g., audio-text and visual-text) to enhance the audiovisual-text alignment, thereby enabling better multimodal representation learning.
Therefore, the overall pre-training objective of our CoAVT consists of three pair-wise losses based on Eq. (\ref{eq6}):
\begin{equation}
    \mathcal{L}_{total} = \mathcal{L}_{AV} + \mathcal{L}_{A} + \mathcal{L}_{V}   
\end{equation}

\subsection{Fine-tuning}
To verify the effectiveness of the learned representations encompassing audio, visual and text, we further fine-tune the CoAVT model for downstream retrieval and classification tasks.
\subsubsection{Fine-tuning for Retrieval}
Video retrieval aims to retrieve the relevant video segment given a free form natural language query.
Unlike most existing video retrieval methods that solely focus on aligning text with visual elements and disregard audio information, our proposed model, empowered with tri-modality encoding ability, enables a holistic exploration of both visual and audio information for text-to-video retrieval.
Moreover, we also consider audio retrieval for downstream evaluation.
During the fine-tuning process, we adhere to the same objectives as pre-training.
Without encoding audio information, existing video retrieval works only focus the matching between text and vision modality.
Benefiting from the tri-modality encoding ability of our model, we fully explore both vision and audio information in the video for text-to-video retrieval.

\subsubsection{Fine-tuning for Audio-Visual Event Classification}
Besides the retrieval task, audio-visual event classification is another challenging task on video understanding, which requires good joint audio-visual representation.
To conduct classification, we apply average pooling on top of the query encoder followed by a randomly initialized linear layer.
Specifically, we fine-tune the model using audio-only data (A), video-only data (V), and audio-visual data (AV) to evaluate the single modal and multi-modal representation quality.

\begin{table*}
\centering
\caption{Comparison with state-of-the-art video retrieval methods on AudioCaps dataset. Inputs refer to video inputs as follows: \textbf{A:} audio spectrogram, \textbf{V:} video frames.
$^{*}$VALOR uses four datasets for pre-training, containing VALOR-1M, WebVid-2.5M, CC14M, and HD\_VILA\_10M.
}
\begin{tabular}{lclcccc}
\toprule
Method                         & Pre-training                                                                        & \#Example             & Modality & R@1  & R@5  & R@10 \\ \hline
\textit{Zero-shot}                      &                                                                                     &                       &          &      &      &      \\
\multirow{2}{*}{AVR~\cite{videcc}}      & \multirow{2}{*}{VideoCC3M}                                                          & \multirow{2}{*}{9.4M} & A        & 8.7  & -    & 37.7 \\
                               &                                                                                     &                       & A+V      & 10.6 & -    & 45.2 \\ \hline
\multirow{2}{*}{CoAVT$_{Baseline}$} & \multirow{6}{*}{AudioSet}                                                           & \multirow{6}{*}{1.4M} & A        & 9.4 & 31.0 & 45.7 \\
                               &                                                                                     &   
                               & A+V      & 10.5 & 33.9 & 48.7 \\
\multirow{2}{*}{CoAVT$_{Vanilla}$} &                                                            &  & A        & 13.0 & 36.1 & 50.9 \\
                               &                                                                                     &                       & A+V      & 13.1 & 36.9 & 51.7 \\
\multirow{2}{*}{CoAVT}          &                                                                                     &                       & A        & 14.1 & 39.3 & 54.7 \\
                               &                                                                                     &                       & A+V      & \textbf{14.3} & \textbf{41.3} & \textbf{57.6} \\ \toprule
\textit{Fine-tuned}                     &                                                                                     &                       &          &      &      &      \\
\multirow{2}{*}{Oncescu et al. 2021}      & \multirow{2}{*}{-}                                                                  & \multirow{2}{*}{-}    & A        & 24.3 & -    & 72.1 \\
                               &                                                                                     &                       & A+V      & 28.1 & -    & 79.0 \\
CLIP4VLA~\cite{clip4vla}                       & AudioSet+HowTo100M                                                                  & 2.5M                  & A        & 28.4 & 60.9 & 76.2 \\
VALOR~\cite{valor}                          & VALOR$^{*}$ & 27.5M                 & A        & 40.1 & 73.9 & 83.1 \\
\multirow{2}{*}{AVR~\cite{videcc}}      & \multirow{2}{*}{VideoCC3M}                                                          & \multirow{2}{*}{9.4M} & A        & 35.5 & -    & 84.5 \\
                               &                                                                                     &                       & A+V      & 43.2 & -    & 88.9 \\ \hline
\multirow{2}{*}{CoAVT$_{Baseline}$} & \multirow{6}{*}{AudioSet}                                                     & \multirow{6}{*}{1.4M} & A        & 33.2 & 67.6 & 80.4 \\
                               &                                                                                     &                       & A+V      & 38.1 & 75.4 & 87.1 \\
\multirow{2}{*}{CoAVT$_{Vanilla}$} &                                                            &  & A        & 36.9 & 71.5 & 83.5 \\
                               &                                                                                     &                       & A+V      & \textbf{45.0} & 79.1 & 89.4 \\
\multirow{2}{*}{CoAVT}          &                                                                                     &                       & A        & 41.8 & 76.5 & 87.8 \\
                               &                                                                                     &                       & A+V      & 44.9 & \textbf{79.3} & \textbf{89.6} \\ \toprule
\end{tabular}
\label{table:audiocaps}
\end{table*}

\section{Experimental Results and Analysis}
\label{exps}
\subsection{Datasets}
To validate the proposed method, we first pre-train the CoAVT model on the large-scale multimodal dataset AudioSet~\cite{audioset}, then fine-tune it for video retrieval, audio-visual event classification and audio-visual retreival tasks on four datasets: AudioCaps~\cite{audiocaps}, Clotho~\cite{clotho}, AudioSet-20K~\cite{audioset} and VGGSound~\cite{vggsound}.
The evaluation metrics are Recall@n (R@n) for retrieval task, and mean average precision (mAP) for classification task.

\subsubsection{Pre-training Dataset}
For pre-training, we use the publicly available large-scale multimodal dataset AudioSet~\cite{audioset}.
It contains over 2 million 10-second YouTube video clips, and each clip is labeled with event labels from a set of 527 distinct labels in a non-exclusive way.
After filtering out those unavailable data, we finally downloaded 1,450,529 audio-video-text pairs for training.
To generate coherent captions from discrete labels, we opt for a simply concatenation without any prompt.

\subsubsection{Fine-tuning Datasets}
We evaluate the pre-trained CoAVT on retrieval and classification benchmarks, including AudioCaps~\cite{audiocaps}, Clotho~\cite{clotho}, AudioSet-20K~\cite{audioset} and VGGSound~\cite{vggsound}.
\begin{itemize}
\item 
\textbf{AudioCaps} is an audio-centric video dataset, whose videos are mainly in event scenarios with duration shorter than 10 seconds from YouTube.
Each training sample contains one caption, while five captions per sample are used in validation and test sets.
We use this dataset for text-video retrieval task on both zero-shot and fine-tuning settings.
After filtering out the videos that are no longer available, we finally obtain 32,747 training, 442 validation, 753 test samples.

\item 
\textbf{Clotho} is an audio-only dataset of described sounds, which are sourced from Freesound platform~\cite{freesound}.
This dataset consists of a development set and evaluation set of 2893 and 1045 audio samples respectively, and every audio sample is accompanied by 5 captions. For fair comparison, we follow~\cite{Oncescu} and treat each of the 5 captions per test audio as a separate query.
We use this dataset to validate the generalization ability of our CoAVT on text-to-audio retrieval.

\item \textbf{AudioSet-20K and VGGSound} \quad
For the audio-visual event classification task, we conduct experiments on AudioSet-20K~\cite{audioset} and VGGSound~\cite{vggsound}.
AudioSet-20K is a subset of AudioSet-2M with a more balanced class distribution. We downloaded 18,063 training and 16,690 evaluation samples.
VGGSound is a collection of 200K 10-second YouTube video clips annotated with 309 classes, and we downloaded 162,567 training and 13,483 test samples.
\end{itemize}

\subsection{Experimental Settings}
In this section, we provide more details of the input pre-processing, model parameters, and hyper-parameter settings during our experiments.

For the audio input, we first downsample the audio waveform to 16000 Hz, then extract 128-dimensional log Mel-filterbank features with a 25ms Hanning window every 10ms, which results in a 1024 × 128 spectrogram.
For the visual input, to lower the computational overhead, we uniformly sample 10 RGB frames from each 10-second video clip (i.e., 1 FPS), and randomly select one frame as the input during training.
During the inference on retrieval task, the same extraction procedure was performed, with the difference that only the central frame was presented to the model.
While during the inference on audio-visual event classification task, we follow~\cite{cavmae} and use a frame aggregation strategy, in which we average the model prediction of each frame as the final model prediction.

By default, all transformer encoder layers are 768-dimensional and have 12 attention heads.
For the joint audio-visual encoder, the audio and visual encoders are 11-layer Transformer and the joint encoder is a single self-attention Transformer layer.
We initialize the joint audio-visual encoder with CAV-MAE pre-trained weights~\cite{cavmae}.
For the query encoder and text encoder, besides the cross-attention layers, both of them share the same parameters, which are initialized using the BERT$_{base}$~\cite{bert} model.
We set the number of queries to 16 as default setting if not specified otherwise.
To improve the training efficiency, we follow FLIP~\cite{flip} to randomly mask out the audio and visual input and removes a large portion of patches during pre-training, with the probability of 0.75 and 0.5 for audio and video respectively as our default settings if not specified otherwise.
While during fine-tuning, we do not apply masking on audio and visual input.
We pre-train the model for 5 epochs using a batch size of 512 on 8 NVIDIA A100 GPUs.
We use AdamW~\cite{adamw} optimizer with $\beta_1=0.9$, $\beta_2=0.98$, and a weight decay of 0.05 for both pre-training and fine-tuning.
We use a cosine learning rate decay with a peak learning rate of 1e-4 and a linear warmup of 2k steps. The minimum learning rate at the second stage is 1e-6.
For the fine-tuning on retrieval task, the learning rate is set to 3e-5 and the model is trained for 15 epochs with a batch size of 128 on 4 NVIDIA A100 GPUs.
For the fine-tuning on event classification task, the learning rate is set to 8e-5 and the model is trained for 15 epochs with a batch 64 on one GPU.

\begin{table*}[]
\centering
\caption{Comparison with state-of-the-art video retrieval methods on AudioCaps dataset. Inputs refer to video inputs as follows: \textbf{A:} audio spectrogram, \textbf{V:} video frames.
$^{*}$VALOR uses four datasets for pre-training, containing VALOR-1M, WebVid-2.5M, CC14M, and HD\_VILA\_10M.
}
\begin{tabular}{lccccccc}
\toprule
\multirow{2}{*}{Method} & \multirow{2}{*}{Pre-training} & \multicolumn{3}{c}{AudioSet-20K (mAP)} & \multicolumn{3}{c}{VGGSound (Acc)} \\ \cline{3-8} 
                        &                               & A           & V           & A+V        & A          & V         & A+V       \\ \hline
\textit{Audio-only Models}       &                               &             &             &            &            &           &           \\
PANNS~\cite{panns}                   & -                             & 27.8        & -           & -          & -          & -         & -         \\
AST~\cite{ast}                     & SL                            & 34.7        & -           & -          & -          & -         & -         \\
SSAST~\cite{ssast}                   & SSL                           & 31.0        & -           & -          & -          & -         & -         \\
MAE-AST~\cite{maeast}                 & SSL                           & 30.6        & -           & -          & -          & -         & -         \\
Audio-MAE~\cite{audiomae}               & SSL                           & 37.1        & -           & -          & -          & -         & -         \\
Chen et al. 2020                   & -                             & -           & -           & -          & 48.8       & -         & -         \\
AudioSlowFast~\cite{audioslowfast}           & -                             & -           & -           & -          & 50.1       & -         & -         \\ \hline
\textit{Audio-Visual Models}     &                               &             &             &            &            &           &           \\
G-Blend$^{*}$~\cite{gblend}                 & -                             & 29.1        & 22.1        & 37.8       & -          & -         & -         \\
\textcolor{gray}{MBT$^{*}$~\cite{mbt}}                     & \textcolor{gray}{SL}                            & \textcolor{gray}{31.3}        & \textcolor{gray}{27.7}        & \textcolor{gray}{43.9}       & \textcolor{gray}{52.3}       & \textcolor{gray}{51.2}      & \textcolor{gray}{64.1}     \\
CAV-MAE~\cite{cavmae}                 & SSL                           & 37.7        & 19.8        & 42.0       & 59.5       & 47.0      & 65.5      \\
MAViL~\cite{mavil}                   & SSL                           & 39.0        & \textbf{22.2}        & 42.5       & 59.9       & \textbf{48.3}      & 63.8      \\ \hline
CoAVT$_{Vanilla}$          & SL                           & 38.9        & 18.0        & \textbf{45.0}       &  60.1        &   46.7        &   66.2        \\ 
CoAVT                    & SL                           & \textbf{42.5}        & 20.6        & 44.7       & \textbf{60.7}       & 48.1      & \textbf{66.4}      \\ \toprule
\end{tabular}
\label{table:classification}
\end{table*}

\subsection{Results}
\subsubsection{Text-to-Video Retrieval}
To demonstrate the effectiveness of our proposed CoAVT model, we first evaluate it for video retrieval on AudioCaps benchmark.
During inference, we follow~\cite{alpro, blip2}, which first select $k=128$ candidates based on the similarity scores, then followed by a re-ranking based on pairwise matching scores.
The results are provided in Table~\ref{table:audiocaps}.
We mainly compare the CoAVT with three tri-modal pre-training methods, including 1) CLIP4VLA~\cite{clip4vla}, which incorporates audio into VL pre-training framework CLIP and trains on both AudioSet and HowTo100M datasets; 2) VALOR~\cite{valor}, which uses three separate encoders and trains on four large-scale datasets, containing 27.5M samples in total; 3) AVR~\cite{videcc}, which uses a dual-stream model but trains with simple contrastive loss between audiovisual and text features on VideoCC3M dataset.
For fair comparison, we build another two baseline models: $Baseline$ model use the same objective as AVR but trains with our AudioSet dataset, while $Vanilla$ model employs a query encoder between the joint audiovisual encoder and text encoder, ignoring the bi-modal audio-text and visual-text alignments, and trains with three objectives on AudioSet.

\begin{table}[]
\centering
\caption{Results on the Clotho dataset for text-audio retrieval. Note this dataset only contains audio information, no visual track.
$^{*}$VALOR uses four datasets for pre-training, containing VALOR-1M, WebVid-2.5M, CC14M, and HD\_VILA\_10M, resulting in total 27.5M samples.
}
\scalebox{0.95}{
\begin{tabular}{lccccc}
\toprule
Method       & Pre-training  & Fine-tuning                                                                            & R@1  & R@5  & R@10 \\ \hline
Oncescu.~\cite{Oncescu}     & -         &      Clotho                                                                       & 9.6  & -    & 40.1 \\
\multirow{2}{*}{LAION~\cite{laion}}         &      & \multicolumn{1}{c}{\multirow{2}{*}{\begin{tabular}[c]{@{}c@{}}Clotho\\ + AudioCaps\end{tabular}}}   & \multirow{2}{*}{12.0} & \multirow{2}{*}{31.6} & \multirow{2}{*}{43.9} \\ 
         &      &  \multicolumn{1}{c}{}      & & & \\
\multirow{2}{*}{AVR~\cite{videcc}} & \multirow{2}{*}{VideoCC3M}          &    -                                                                   & 3.0  & -    & 17.5 \\
     &          &     Clotho                                                                   & 12.6 & -    & 45.4 \\
VALOR~\cite{valor}        &VALOR* &   Clotho  &  17.5 & 42.7 & 55.3 \\ \hline

\multirow{4}{*}{CoAVT}   & \multirow{4}{*}{AudioSet}    
         &      -        & 4.7  & 13.8 & 20.3 \\
         &      &    Clotho                                                                         & 13.7 & 33.8 & 44.9 \\ 
         &      & \multicolumn{1}{c}{\multirow{2}{*}{\begin{tabular}[c]{@{}c@{}}Clotho\\ + AudioCaps\end{tabular}}}   & \multirow{2}{*}{16.4} & \multirow{2}{*}{37.9} & \multirow{2}{*}{50.0} \\ 
         &      &  \multicolumn{1}{c}{}      & & & \\
\toprule
\end{tabular}
}
\label{table:clotho}
\end{table}

Table~\ref{table:audiocaps} compares the video retrieval performances of different approaches on the AudioCaps corpus.
The results show that the proposed CoAVT model achieves state-of-the-art performance with significant improvement over existing methods in both zero-shot and fine-tuned settings, either with only audio representations or joint audiovisual representations.
When compared to the strong baseline AVR model, our $Baseline$ model obtains similar performance under the zero-shot setting (e.g., 10.6/45.3 $\rightarrow$ 10.5/48.7), despite that it is pre-trained on a much larger video-text dataset~\cite{videcc}, which contains more than 9M samples.
Furthermore, our $Vanilla$ model, which employs the query encoder to address the modality gap, outperforms the AVR model (e.g., 10.6/45.3 $\rightarrow$ 14.3/57.6), which demonstrates the effectiveness of the query encoder.
Finally, when our final model is further equipped with audio-text and video-text correspondences during pre-training, we obtain the state-of-the-art retrieval results.
This indicates that our model is capable of learning more effective semantic representations across audio, vision and text modalities.

We further evaluate our model by conducting audio-only retrieval on the Clotho dataset.
As the results shown in Table~\ref{table:clotho}, our CoAVT model outperforms the AVR model, but falls short of the VALOR model.
It should be noted that VALOR model uses a significantly larger dataset during pre-training, with up to 27.5M samples, compared to the 1.4M samples used in the pre-training of our CoAVT model.
Despite this difference in pre-training sample size, the CoAVT model still demonstrates competitive performance, underscoring its efficiency and potential.

\subsubsection{Audio-Visual Event Classification}
To further validate whether our CoAVT model could learn a good joint audio-visual representation, we conduct audio-visual event classification experiments.
Table~\ref{table:classification} summarizes the performance comparison on AudioSet and VGGSound datasets.
We report accuracy for fine-tuning using the audio(A), video(V) and joint audio-visual representation (A+V).
Our CoAVT sets new state-of-the-art performance on audio-only and audio-visual classification on both AudioSet-20K and VGGSound datasets.
As shown in the table, our $Vanilla$ model already outperforms recent competitive models, such as CAV-MAE~\cite{cavmae} and MAViL~\cite{mavil} by a large margin (e.g., 42.0/42.5 $\rightarrow$ 45.0 and 65.5/63.8 $\rightarrow$ 66.2) on audio-visual based event classification.
Furthermore, when we include the audio-text and visual-text bi-modal information during pre-training, our CoAVT also surpasses these models on audio-only based event classification (e.g., 37.7/39.0 $\rightarrow$ 42.5 and 59.5/59.9 $\rightarrow$ 60.7).
This indicates that our CoAVT model not only captures a good  joint audio-visual representation, but also effectively learns audio-only representation.

These results on video retrieval and audio-visual event classification validate the effectiveness of our CoAVT model in handling multimodal understanding tasks and highlight its potential for improving performance in both audio-visual and audio-only scenarios. 
The inclusion of bi-modal audio-text and visual-text bi-modal correspondences during pre-training provides additional benefits, indicating that the integration of bi-modal information can lead to more robust and versatile multimodal representations.

\begin{table*}[]
\centering
\caption{Visual-to-audio retrieval results on the subset of AudioSet and VGGSound.}
\begin{tabular}{lccccccc}
\toprule
\multirow{2}{*}{Method} & \multirow{2}{*}{Pre-training}                  & \multicolumn{3}{c}{AudioSet} & \multicolumn{3}{c}{VGGSound} \\ \cline{3-8} 
                        &                                                & R@1      & R@5     & R@10    & R@1      & R@5     & R@10    \\ \hline
\multicolumn{8}{l}{\textit{Visual-to-Audio Retrieval}}                                                                                         \\
CAV-MAE~\cite{cavmae}                 & \multirow{3}{*}{AudioSet}                      & 18.8     & 39.5    & 50.1    & 14.8     & 34.2    & 44.0    \\
CoAVT$_{Vanilla}$            &                                                & 27.9     & 55.1    & 65.4    & 30.9     & 61.8    & 72.4    \\
CoAVT                     &                                                & 32.1     & 61.5    & 72.4    & 40.9     & 71.2    & 81.8    \\ \hline
\multicolumn{8}{l}{\textit{Visual-to-Audio Retrieval}}                                                                                          \\
CAV-MAE~\cite{cavmae}                 & \multirow{3}{*}{AudiosSet} & 15.1     & 34.0    & 43.0    & 12.8     & 30.4    & 40.3    \\
CoAVT$_{Vanilla}$            &                            & 22.7     & 48.3    & 57.6    & 29.4     & 56.3    & 67.5    \\
CoAVT                     &                           & 33.1     & 63.6    & 73.0    & 36.6     & 66.1    & 78.1    \\ \toprule
\end{tabular}
\label{table:AVretrieval}
\end{table*}

\subsubsection{Audio-Visual Retrieval}
In this section, we further investigate whether our CoAVT model also learns a well-coordinated representation that captures audio-visual correspondence via text for audio-visual retrieval task.
For fair comparison, we use the same subset of AudioSet and VGGSound evaluation set as in ~\cite{cavmae}, which includes 1,725 and 1,545 audio-visual samples from the AudioSet and VGGSound evaluation sets, respectively.
Specifically, we input audio and image to the model in two independent forward passes and take the mean-pooled query encoder outputs as audio and visual representation, respectively.
We then calculate the retrieval recall at rank 1,5 and 10 (R@1, R@5, R@10) based on the cosine similarity of the audio and visual representation. 
The results are shown in Table~\ref{table:AVretrieval}.
As demonstrated in the table, compared to CAV-MAE, which learns the joint audio-visual representation through self-supervised learning, our model could leverage additional textual information as a bridge and achieves significantly better audio-visual retrieval performance.
This result indicates that our CoAVT model is not only capable of learning high-quality joint audio-visual representations, but also excels in coordinating these representations to capture audio-visual correspondences effectively.
The use of text as a bridge between audio and visual modalities enhances the model's performance in the audio-visual retrieval task.

\begin{table}[]
\centering
\caption{Effects of the number of learnable queries of the query encoder during pre-training.}
\begin{tabular}{lcccc}
\toprule
\# query            & \multicolumn{1}{l}{Modality} & \multicolumn{1}{l}{R@1} & \multicolumn{1}{l}{R@5} & \multicolumn{1}{l}{R@10} \\ \hline
\multirow{2}{*}{8}  & A                            & 11.4                    & 33.4                    & 47.5                     \\
                    & A+V                          & 12.9                    & 35.9                    & 50.3                     \\ \hline
\multirow{2}{*}{16} & A                            & 13.0                    & 36.1                    & 50.9                     \\
                    & A+V                          & 13.1                    & 36.9                    & 51.7                     \\ \hline
\multirow{2}{*}{32} & A                            & 10.8                    & 31.7                    & 46.6                     \\
                    & A+V                          & 10.4                    & 33.6                    & 49.1                     \\ \toprule
\end{tabular}
\label{table:ablation_query}
\end{table}

\begin{table}[]
\centering
\caption{Effects of the number of learnable queries of the query encoder during pre-training.}
\begin{tabular}{lcccc}
\toprule
Masking ratio                                                              & Modality & R@1  & R@5  & R@10 \\ \hline
\multirow{2}{*}{no mask}                                                   
                                                                           & A        & 12.4 & 37.4 & 52.5 \\
                                                                           & A+V      & 13.3 & 38.8 & 55.4 \\ \hline
\multirow{2}{*}{\begin{tabular}[c]{@{}l@{}}ma=0.75\\ mv=0.75\end{tabular}} 
                                                                           & A        & 13.5 & 41.0 & 55.9 \\
                                                                           & A+V      & 14.3 & 41.1 & 57.3 \\ \hline
\multirow{2}{*}{\begin{tabular}[c]{@{}l@{}}ma=0.75\\ mv=0.5\end{tabular}}  
                                                                           & A        & 14.1 & 39.3 & 54.7 \\
                                                                           & A+V      & 14.3 & 41.3 & 57.6 \\ \hline
\multirow{2}{*}{\begin{tabular}[c]{@{}l@{}}ma=0.5\\ mv=0.75\end{tabular}}  
                                                                           & A        & 13.1 & 39.3 & 54.5 \\
                                                                           & A+V      & 13.7 & 40.0 & 56.0 \\ \toprule
\end{tabular}
\label{table:ablation_masking}
\end{table}

\subsection{Ablation Study}
In this section, we conduct ablation experiments to further understand the contributions of different components of our CoAVT model.
The experiments are conducted on the zero-shot video retrieval task using AudioCaps dataset.


\subsubsection{Effect of Number of Queries}

The learnable query embeddings of the query encoder plays an important role in extracting the audio-visual features that are informative of the corresponding text.
To investigate the effect of the number of queries, we vary the number from 8 to 32 during pre-training based on our $Vanilla$ model.
The results are shown in Table~\ref{table:ablation_query}. 
From these results, it becomes clear that the number of queries has a significant impact on the model's performance. 
The model achieves the best performance when the number of queries is set to 16. 
Interestingly, when the number of queries is increased from 16 to 32, the performance degrades dramatically.
This might be due to that increasing the number of queries will introduce more noise and redundancy in the model. 
These additional queries may capture irrelevant or redundant information, which can interfere with the learning process and hinder the model's ability to extract meaningful multimodal representations, ultimately leading to a decrease in performance.

\subsubsection{Effect of Masking Ratio}
Masking is a crucial technique used during pre-training, which involves randomly hiding a portion of the input data and prompting the model to predict the masked data based on the context provided by the unmasked data. 
This encourages the model to learn robust and generalizable representations of the data.
In our CoAVT model, we perform both audio and video masking during pre-training. 
To investigate the impact of masking, we conduct an ablation study by varying the masking ratio.
As shown in Table~\ref{table:ablation_masking}, the model yields best performance when the masking ratio of audio is 0.75 and the masking ratio of video is 0.5.
Compared to the models that do not use any masking, the model with masking not only makes the pre-training process more efficient but also improves the generalization ability of the model.

\subsubsection{Effect of Audio-Text and Visual-Text Alignment}
To enhance the multimodal representation learning, we introduce the audio-text and visual-text bi-modal alignments in addition to the base audiovisual-text tri-modal alignment in the CoAVT model.
These alignments aim to capture the rich semantic correspondences between different modalities, thus improving the model's ability to understand and integrate multimodal information.
In order to evaluate the contribution of audio-text and visual-text alignment separately, we conduct an ablation study where we remove these alignments one by one during pre-training.
The results are shown in Table~\ref{table:objective}.
When we remove the visual-text alignment, the overall performance drops, especially in the video-only based retrieval task. 
This indicates that the visual-text alignment plays a crucial role in capturing the semantic correspondences between visual and textual data, which is particularly important for tasks that rely heavily on visual information.
When we further remove the audio-text alignment, the performance of retrieval tasks degrades accordingly. 
This suggests that the audio-text alignment also contributes significantly to the model's performance, helping to establish meaningful correspondences between audio and textual data.

Overall, these results reveal that the bi-modal correspondences help the model capture rich semantic representations between modalities, leading to better performance in downstream tasks. This underscores the importance of incorporating bi-modal alignments in multimodal representation learning models.


\begin{figure*}
\centering
\includegraphics[scale=0.7]{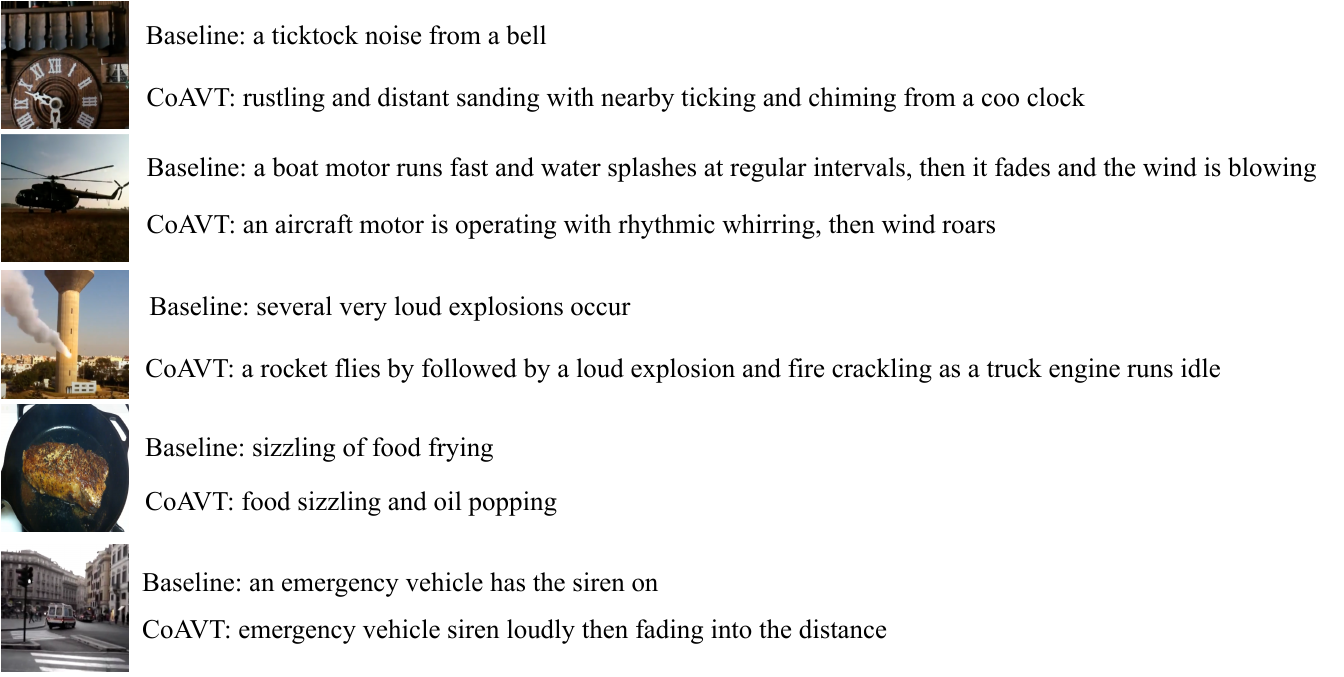}
\caption{Qualitative results of video-to-text retrieval on AudioCaps.}
\label{fig:examples}
\end{figure*}

\begin{table}[]
\centering
\caption{Effects of the correspondence of audio-text and visual-text.}
\begin{tabular}{lcccc}
\toprule
Pre-training loss                      & Modality & R@1  & R5   & R@10 \\ \hline
\multirow{3}{*}{$\mathcal{L}_{total}$}       & V        & 5.6  & 19.9 & 30.9 \\
                            & A        & 12.4 & 37.4 & 52.5 \\
                            & A+V      & 13.3 & 38.8 & 55.4 \\ \hline
\multirow{3}{*}{\quad  -$\mathcal{L}_V$} & V        & 3.0  & 13.7 & 23.7 \\
                            & A        & 13.1 & 37.4 & 52.5 \\
                            & A+V      & 13.6 & 37.9 & 53.6 \\ \hline
\multirow{3}{*}{\quad -$\mathcal{L}_V$-$\mathcal{L}_A$} & V        & 3.7  & 14.5 & 24.4 \\
                            & A        & 13.0 & 36.1 & 50.9 \\
                            & A+V      & 13.1 & 36.9 & 51.7 \\ \toprule
\end{tabular}
\label{table:objective}
\end{table}

\subsubsection{Qualitative Analysis}
In Figure~\ref{fig:examples}, we show some qualitative results of our CoAVT and the $baseline$ model on video-to-text retrieval task.
Specifically, we present the query videos with matched texts.
Our CoAVT model demonstrates a stronger cross-modal video-text understanding capability compared to the baseline model.
For example, consider the second video in the figure. While the baseline model returns a more generic or less accurate description, the CoAVT model returns the text ``an aircraft motor", which is a more specific and accurate description of the audiovisual content in the video.
These qualitative results suggest the superiority of our CoAVT model over the baseline model in the downstream task. 
This reinforces the effectiveness of our approach, which leverages bi-modal and tri-modal alignments for enhanced multimodal representation learning. 
The results also highlight the model's ability to capture fine-grained semantic correspondences between video and text, which is crucial for various downstream tasks.

\section{Conclusion}
\label{conclusion}
This study aimed to develop a new tri-modal audio-visual-text pre-training method for multimdodal processing.
The proposed CoAVT model employs a joint audio-visual encoder to handle audio and visual input simultaneously, and a text encoder for textual input.
This dual model has two subsystems, one is for non-verbal information (e.g., audio and video), and one is for verbal information (e.g., text), similar to human cognition mechanisms.
CoAVT introduces a query encoder to bridge the modality gap, thereby enabling better alignment of the representations from different modality.
It further establishes the bi-modal alignments upon the base audiovisual-text tri-modal alignment, which explicitly correlate audio, video and text modalities.
The effectiveness of the proposed pre-training method was validated on three downstream tasks, including video retrieval, audio-visual event classification, and audio-visual retrieval tasks.
Extensive experimental results clearly demonstrated, as a unified audio-visual-text model, its consistent superiority for multimodal understanding.


\bibliographystyle{IEEEtran}
\bibliography{references}

\end{document}